Title
# Ultra-broadband On-chip Twisted Light Emitter

Authors

Zhenwei Xie[1]†, Ting Lei[1]†*, Fan Li[2], Haodong Qiu[3], Zecen Zhang[3], Hong Wang[3], Changjun Min[1], Luping Du[1], Zhaohui Li[2]*, Xiaocong Yuan[1]*

Affiliations

[1]Nanophotonics Research Center, Shenzhen University & Key Laboratory of Optoelectronic Devices and Systems of Ministry of Education and Guangdong Province, College of Optoelectronic Engineering, Shenzhen University, Shenzhen, 518060, Guangdong, China.

[2]State Key Laboratory of Optoelectronic Materials and Technologies and School of Electronics and Information Technology, Sun Yat-sen University, Guangzhou 510275, China.

[3] School of Electrical and Electronic Engineering, Nanyang Technological University, Singapore 639798.

*Corresponding authors. E-mail: xcyuan@szu.edu.cn (X. C. Yuan); li_zhaohui@hotmail.com (Z. Li); leiting@szu.edu.cn (T. Lei).

† Zhenwei Xie and Ting Lei contributed equally to this work.

## Abstract

On-chip twisted light emitters are essential components for orbital angular momentum (OAM) communication devices, which could address the growing demand for high-capacity communication systems by providing an additional degree of freedom for wavelength/frequency division multiplexing (WDM/FDM). Although whispering gallery mode enabled OAM emitters have been shown to possess some advantages, such as being compact and phase accurate, their inherent narrow bandwidth prevents them from being compatible with WDM/FDM techniques. Here, we demonstrate an ultra-broadband multiplexed OAM emitter that utilizes a novel joint path-resonance phase control concept. The emitter has a micron sized radius and nanometer sized features. Coaxial OAM beams are emitted across the entire telecommunication band from 1450 to 1650 nm. We applied the emitter for OAM communication with a data rate of 1.2 Tbit/s assisted by 30-channel optical frequency combs (OFC). The emitter provides a new solution to further increase of the capacity in the OFC communication scenario.

## MAIN TEXT

## Introduction

Twisted light with a helical wavefront, carrying orbital angular momentum (OAM) (*1*), has a phase profile of $\exp(im\varphi)$, where $m$ is the topological charge indicating the mode order of the OAM, and $\varphi$ is the azimuthal angle (*2*). OAM beams are widely used in numerous applications, including particle trapping (*3, 4*), quantum memory (*5*), on-chip optical encoding (*6*), optical metrology, and microphotography (*7, 8*). Recently, OAM has attracted considerable attentions for use in mode division multiplexing in free space and fiber applications for both classical and quantum communication systems (*9-14*); the infinite number of orthogonal states of OAM provide an additional degree of freedom to the conventional multiplexing techniques (*15*). On-chip optical interconnects are increasingly replacing wires owing to the significant requirements for massive data link capacities in high performance computers and warehouse-scale datacenters, OAM orientated communication via on-chip optical interconnects could provide new opportunities for additional multiplexing to address the need for high capacity, low power communications technology that is highly compatible with existing wavelength/frequency division multiplexing (WDM/FDM) systems. These possibilities motivated pioneering research into integrated silicon OAM emitters and III-V OAM micro-lasers based on whispering gallery-mode (WGM) resonators (*16-18*). However, the intrinsic high quality factor (Q) of the WGM limits the emission bandwidth and significantly hinders the compatibility of these devices for WDM/FDM compatible applications (*19*). This presents a significant obstacle for future on-chip communication, since on-chip optical frequency combs (OFCs) (*20-24*) enabled broadband emitters are set to dominate the next generation of multiplexing technologies (*25, 26*). Therefore, OAM emitters with broadband capability are required that are able to convert OFCs to twisted light for further expansion of communication capacity.

In this work, we propose and demonstrate a silicon-based multiplexed OAM emitter with ultra-broad bandwidth in the telecommunication band (from 1450 to 1650 nm); this is achieved via the joint phase control of the optical path and the local resonances in subwavelength structures. The working principle is based on synthesis of a series of nano-sized cavities with low Q resonance, enabling accurate phase control as well as short response time. Figure 1A shows a schematic of the multiplexed OAM emitter for broadband input signals with multiple frequencies from $f_1$ to $f_n$. The emitter has a circular shape containing subwavelength structures and is connected to two single-mode waveguides. The OFCs signals are input from the left- or right-hand-side of the waveguide (left or right arm of the device), oscillate in the device, and emit vertically into free space in the form of the OAM modes with state numbers of −1 or +1, respectively. Therefore, all the frequency channels are simultaneously multiplexed by the OAM through the emitter.

**Results**

Our OAM emitter was fabricated on a standard silicon on insulator wafer with a silicon layer of 220 nm and a buried oxide layer of 2 μm. The radius of the device was chosen to be 1.2 μm and inner resonance structures were 100s nm in size (Fig. S1A) (*27*). Figure 1B shows the detailed structures and the phase modulation scheme of the OAM emitter. The total phase modulation originates both the propagation phase delay and the resonance phase delay from the local subwavelength structure. Thus the phase modulation for the emitter can be expressed as

$$\phi(r,\varphi) = \phi_1(r,\varphi) + \phi_2(r,\varphi) = \int_{P(R,\pi)}^{P(r,\varphi)} \frac{2\pi N(r,\varphi)}{\lambda}dl + \frac{8\pi N^4(r,\varphi)V}{D\lambda^3} + \phi', \tag{1}$$

where $r, \varphi$ are the polar coordinates (origin is the center of the device), $\phi$ is the phase modulation of the emitter, $\phi_1$ and $\phi_2$ are the phase modulations induced by the propagation and localized resonance, respectively. $\phi_1$ is calculated from the integral along the propagation path, where $P(R,\pi)$ is the link point for the device and its left arm, $P(r,\varphi)$ is the local point in the device where the light is emitted out of plane, $N(r,\varphi)$ is the refractive index distribution of the device determined by the detailed substructures. The estimated propagation delay lies somewhere in the range of 0 to 6π, corresponding to a response time less than 15fs. $\phi_2$ is derived with respect to the local refractive index distribution and structure volume, $V$ is the volume of the local cavity, and $D$ is the mode degeneracy, $\phi'$ is the difference between the phase delay and the group delay of the local resonance, and its value lies between -π~π. The substructures of the designed devices have volumes ranging from $1.6\times10^{-4}\lambda^3$ to $4.7\times10^{-3}\lambda^3$. The calculated phase delay is sufficient for a 2π phase modulation while the corresponding Q of the local cavity is below 5. Therefore, the designed device is equipped with fully phase control and ultrafast time response properties. According to the time-bandwidth product relation, the emitter should also have a broadband spectral domain. Considering that light coming from the left arm yields the $m_1$ order OAM mode, and light coming from the right-hand-side yields the $m_2$ order OAM mode, we obtain the following set of equations:

$$\begin{cases} \int_{P(R,\pi)}^{P(r,\varphi)} \frac{2\pi N(r,\varphi)}{\lambda} dl + \frac{8\pi N^4(r,\varphi)V}{D\lambda^3} + \phi_L' = m_1\varphi; \\ \int_{P(R,0)}^{P(r,\varphi)} \frac{2\pi N(r,\varphi)}{\lambda} dl + \frac{8\pi N^4(r,\varphi)V}{D\lambda^3} + \phi_R' = m_2\varphi. \end{cases} \quad (2)$$

The solutions for $N(r,\varphi)$ in Eq. (2) determine the detailed substructures of the device. However, the solution for Eq. (2) cannot be derived analytically. Here, we utilized a global optimization algorithm to solve the equation set (Supplementary Materials). For the proof of concept demonstration, we chose parameters $m_1$ and $m_2$ to be +1 and -1 to design a multiplexed emitter for OAM modes with topological orders of +1 and -1. The optimized device with detailed substructures is shown in Fig. 1B; the insert shows an SEM image of the fabricated device.

To further investigate the device's high speed response, we conducted finite-difference time-domain (FDTD) simulations in which we sent a 40 fs signal with a wavelength between 1300 and 1900 nm from the right arm and then measured the response at eight monitor points (see Fig. 2A). The points A1 (0.8 μm, π/2), B1 (0.8 μm, 3π/2), C1 (0.8 μm, π), and D1 (0.8 μm, 0) are all in the plane of the device, while A2, B2, C2, and D2 are the corresponding monitoring points of A1, B1, C1, and D1 but shifted upwards by 600 nm. Since A2 and B2 as well as C2 and D2 are centrally symmetric points in the emitted OAM state of 1, the phase difference between them should be π. From the simulated time domain profiles (Fig. 2B) we deduced that the time responses for points A1 and B1 were almost synchronized, which indicates that the propagation delay for these two points was almost the same. Because of the local resonance, the phase difference between the field radiated from points A2 and B2 was π (±5%) from 38 to 55 fs, as predicted for the OAM state of 1. In this case, the majority of the contribution to the overall phase was from the local resonance. For points C1 and D1, the propagation time delay difference was about 15 fs. The phase difference between points C2 and D2 was adjusted to π (±5%) from 45 to 65 fs via the local resonance. The total phase difference consisted of both propagating phase and resonance phase contributions. These results verify not only the design principle, but also the ultra-fast response of the proposed device. Figure 2C and D show the FDTD simulated time response of the emitted OAM modes with state numbers of +1 and −1 at locations 600 nm above the surface. The total time response was less than 40 fs for the OAM emissions.

Figure 3A shows the FDTD simulation for the transmission efficiency for the OAM emitter in the wavelength range from 1300 to 1900 nm. The OAM +1 mode and −1 mode had 3 dB emission bandwidths that were 412 nm in width (1400 to 1812 nm) and 447 nm in width (1300 to 1747 nm), respectively. To experimentally measure the emission bandwidth, the transmission efficiency was characterized using a tunable laser (Fig. S10) (*27*). Figure 3B shows the emission efficiency of the +1 and −1 mode from 1450 to 1650 nm. The maximum emission efficiency reached 35% at 1550 nm with fluctuations of less than 5 dB over the entire wavelength range of 200 nm. Figure 3C shows the FDTD simulation for emission of the −1 and +1 OAM modes at 1550 nm. The simulated intensity pattern displays a hollow shape, while the phase profiles indicate counterclockwise and clockwise spiral phase. Figure 3B shows the experimentally measured intensity and interference patterns from 1450 to 1650 nm at intervals of 50 nm. The sign of the topological charge is indicated by the chirality of the interference pattern. The measured mode purity ranged from 90% (at 1650 nm) to 97% (at 1550 nm) for the +1 OAM mode and from 84% (at 1650 nm) to 95% (at 1550 nm) for the −1 OAM mode (Fig. S5) (*27*). Although the intensity profiles show asymmetry at some wavelengths, the relatively small phase deviations still ensure the high mode purity of the OAM (see Fig. S4) (*27*). This demonstrates that the generated OAM modes possess the desired phase distribution across the entire 200 nm-wide telecommunication band.

As a proof of concept, we demonstrated the use of the proposed multiplexed OAM emitter for high capacity communication applications by testing it with a commercial FDM system. The FDM module generated a set of 30 frequency combs with 0.2 nm-wide channel intervals. Each frequency channel was loaded with 20 Gbit/s quadrature phase shift keying signals; thus, the total data rate for the emitter associated with the two multiplexed OAM modes was 1.2 Tbit/s. Figure 4A shows the measured bit error rate (BER), and the insert shows the corresponding signal constellation. For all channels, the BER was below the hard decision forward error correction limit of $3.8 \times 10^{-3}$. The BER curves for the back-to-back test case and the emissions of the OAM modes under different received powers were also measured (Fig. 4B).

**Discussion**

In conclusion, we demonstrated a multiplexed on-chip OAM emitter with an ultra-broadband response between 1450 and 1650 nm. Emission efficiencies as high as 35% were achieved, and the mode purity exceeded 97%. Since we utilized standard integrated circuit techniques, the proposed device is suitable for mass production by the microelectronics industry. In conjunction with integrated OFCs [24], the proposed OAM multiplexed emitter paves the way for large capacity chip-to-chip optical interconnects. More importantly, the proposed joint path resonance phase control will inspire more in-depth fundamental research on manipulating and mapping phase structures for general broadband photonic devices.

**Acknowledgments**
**General**: We thank X. Wang, K. Zhang and F. Dong (National Center for Nanoscience and Technology, Beijing, China), for the discussions of the device manufacture.

**Funding:** This work is supported by the National Natural Science Foundation of China (Grant Nos. 61138003, 61427819, 61490712, 61405121, 11604218, 61525502, 61435006, 61490715), the Science and Technology Innovation Commission of Shenzhen (Grant No. KQCS2015032416183980), the Fundamental Research Foundation of Shenzhen (Grant No. JCYJ20140418091413543), the Natural Science Foundation of SZU (Grant Nos. 000011, 000075). X.Y. appreciates the support given by the leading talents of Guangdong province program No. 00201505. T. Lei and L. Du acknowledge the Support by Shenzhen Peacock Plan No. KQTD2015071016560101, KQJSCX20160226193555889. H. Qiu, Z. Zhang, and H. Wang acknowledge the Support by Nation Research Foundation of Singapore (NRF-CRP12-2013-04).
**Author contributions:** X.Y. and T.L. developed the concept presented in this paper. Z. X. carried out the analytical and numerical modeling and design the device. Q. H., Z. Z. and H. W. fabricated the device. Z. X., T. L. and F. L. conducted the measurements. X.Y., Z. L. and T.L. supervised the entire project. Z. X., T. L., L. D. and C. M. wrote the manuscript. All authors discussed the results and commented on the article.
**Competing interests:** The authors declare no competing financial interests.

**Data and materials availability:** All data needed to evaluate the conclusions in the paper are present in the paper and the supporting information

**Figures and Tables**

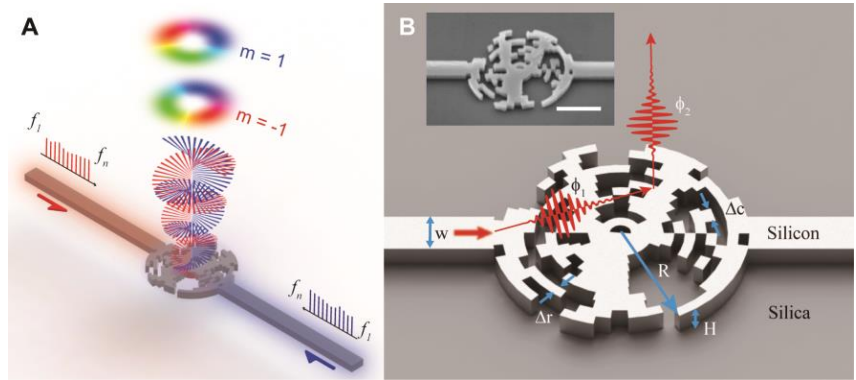

**Fig. 1**. **Schematically illustrated principle and structure design of the broadband multiplexed orbital angular momentum (OAM) emitter.** (**A**) Schematic illustration of the multiplexed OAM emitter operation. (**B**) The details of the structure design. R denotes the radius of the device (R=1.2 μm), H the height of the device (H=220 nm), and W the width of the waveguide (W=440 nm). The red arrow indicates light arriving from the left arm. The two red pulses denote the propagating phase $\phi_1$ and the resonance phase $\phi_2$. The insert shows a scanning electron microscopy (SEM) image of the fabricated OAM emitter with a scale bar of 1 μm.

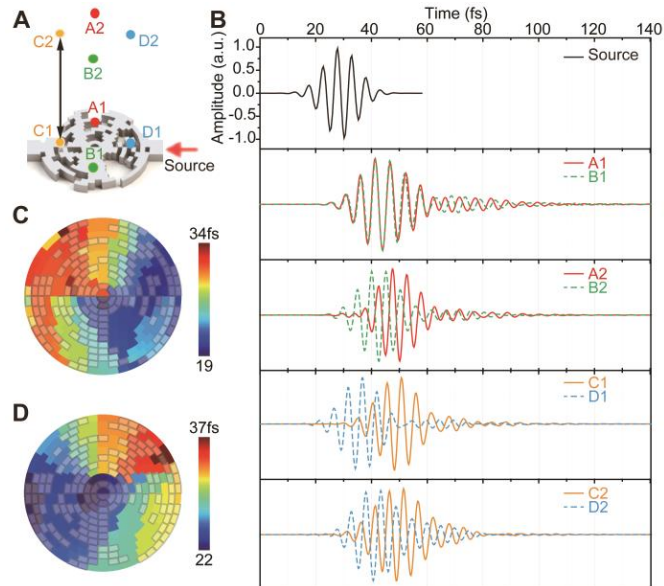

**Fig. 2**. **High speed response of the OAM emitter.** (**A**) The locations of the 4 in-plane monitoring points (A1, B1, C1, and D1) and 4 out-of-plane monitoring points (A2, B2, C2, and D2, which are all located 600 nm above the device) for the OAM emitter time response characterization. (**B**) The finite-difference time-domain (FDTD) simulation results of the time response for the 8 points detailed in panel (A) for the OAM mode +1 emission. Panels (**C**) and (**D**) show the time response of the +1 and −1 OAM modes (group delay), respectively, for each pixel located 600 nm above the surface.

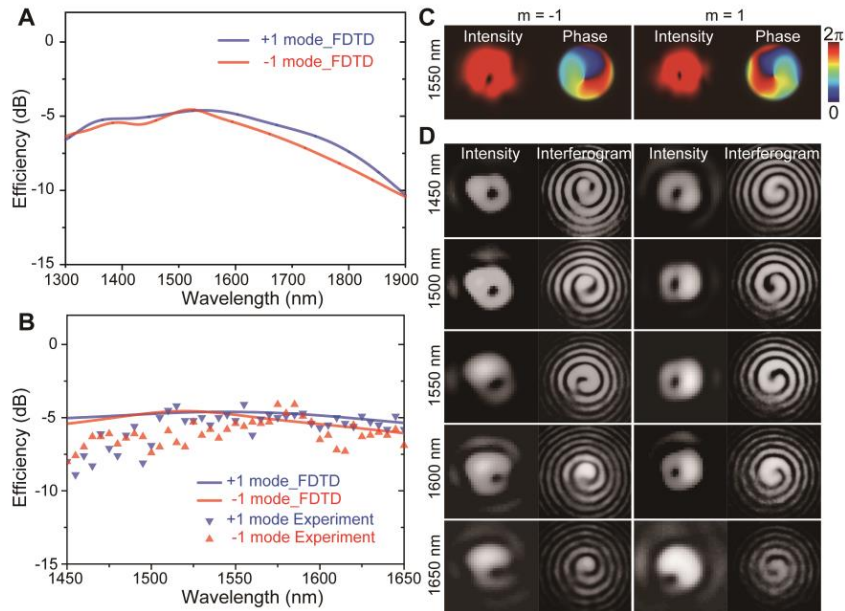

**Fig. 3. Broadband characteristics of the multiplexed OAM emitter.** (**A**) FDTD simulation of the OAM emission efficiency for the wavelength range from 1300 to 1900 nm. (**B**) Comparison between the experimental characterization and FDTD simulation results for the emission of the OAM modes +1 and −1 between 1450 and 1650 nm. (**C**) FDTD simulated near-field intensity (600 nm above the device) and phase profiles for the emission of the −1 and 1 OAM modes at 1550 nm. (**D**) Measured far-field intensity distribution and interference patterns for the emission of the −1 and 1 OAM modes at wavelengths of 1450, 1500, 1550, 1600, and 1650 nm.

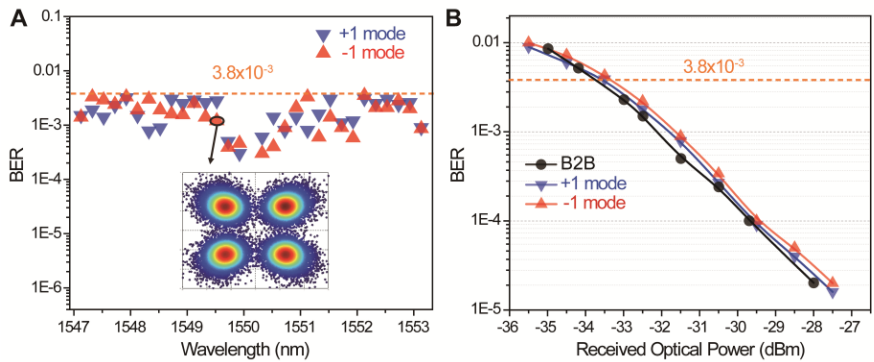

**Fig. 4. Quadrature phase shift keying communication multiplexing results for 30 frequency combs and two OAM modes.** (**A**) The measured bit error rates (BERs) of the +1 OAM mode (blue triangles) and −1 OAM (red triangles). The BERs of all channels are below the hard decision forward error correction limit of $3.8 \times 10^{-3}$. The insert shows the corresponding signal constellations when the BER is $1.0 \times 10^{-3}$. (**B**) BER measurements for back-to-back (B2B) test case, +1 and -1 modes emission at 1550 nm.

**Supplementary Materials:**

Materials and Methods

Supplementary Text

Figures S1-S14

References (*28-29*)

# Supplementary Materials for

## Ultra-broadband On-chip Twisted Light Emitter

Zhenwei Xie, Ting Lei[*], Fan Li, Haodong Qiu, Zecen Zhang, Hong Wang, Changjun Min, Luping Du, Zhaohui Li[*], Xiaocong Yuan[*]

correspondence to:  xcyuan@szu.edu.cn (X. C. Yuan); li_zhaohui@hotmail.com (Z. Li); leiting@szu.edu.cn (T. Lei).

**This PDF file includes:**

    Materials and Methods
    Supplementary Text
    Figs. S1 to S14
    Full Reference List

## Materials and Methods

Fabrication process of the orbital angular momentum (OAM) emitter

As shown in Fig. S1A, the photonic integrated circuit was fabricated on a silicon on insulator wafer with a 220 nm-thick silicon layer on top of a 2 μm-thick buried oxide layer. The layout patterns were defined using a 100 kV electron-beam lithography system (Raith EBPG 5200), while a proximity effect correction was performed using the Layout BEAMER computational software package. A positive e-beam resist (ZEP520A) was employed owing to its high dry etch resistance and high resolution. The photonic structures were then transferred to the silicon device layer via a deep reactive-ion etching process, followed by resist removal with dimethylacetamide.

Global optimization algorithm

To obtain the solution of Eq. (5) in the main article, a global optimization algorithm was proposed and performed. Optimization algorithms are widely utilized for the design of photonic structures. There are several popular algorithms, such as the simulated annealing, genetic, greedy, and ant colony algorithms. Recently, an inverse design algorithm and a direct binary search algorithm have been proposed and implemented for nanophotonic design (*28, 29*). Theoretically, simulated annealing should be very suited to photonic structure design since this method is capable of encompassing the global maximum; however, it is very time consuming. The greedy algorithm, direct binary search algorithm, and genetic algorithm are all more efficient, but they also tend to find the local optimization instead of the global optimization. Therefore, there is a tradeoff between efficiency and performance when choosing a suitable optimization algorithm. Here, we proposed a global optimization method that combines the annealing algorithm and the genetic algorithm associated with three dimensional finite-difference time-domain (FDTD) simulations. Further, to simplify the optimization, we used the inverse process to optimize the structure design, which means that in the optimization process the device was used to couple the two different OAMs to either the left or the right arm. The figure of merit (FOM) for the optimization was defined as the average coupling efficiency for the two OAM modes.

As shown in Fig. S2, the design area was separated into 288 pixels, where each pixel was either silicon or air. The specific optimization process was as follows:

1. Initial guess for the structure design. The initial temperature was set high enough to ensure a sufficient search space. Since the optimization efficiency is dramatically increased by a good initial guess, we chose for the initial guess to be a hologram for both OAM modes, which was then transformed into a binary material distribution.
2. Monitoring the FOM. If the FOM improved, then the temperature was reduced, and the best so far (BSF) and the second BSF (SBSF) values were stored. However, if the FOM got worse, then the SBSF was replaced by chance and a new guess was obtained by randomly mixing the BSF and SBSF.
3. Iteration of the above process until the FOM reached its maximum or the whole iteration exceeded the maximum iteration limit.

Characterizing the intensity and phase profiles and the emission efficiency measurements

Fig. S3 shows the experimental setup for the characterization of the generated phase of the OAM modes and intensity profiles as well as for the emission bandwidth tests. The phase profiles were measured with a classical interferometer approach. After passing a polarization controller, the light from a laser that was tunable across the entire telecommunication band (from 1450 to 1650 nm) was split into two branches by a 3 dB optical coupler. One branch was passed to a variable optical attenuator (2–60 dB) and a collimator to produce an expanded Gaussian beam as the reference beam. A half wave plate (HWP) was utilized to ensure a consistent polarization between the emission beam and the reference beam. The other branch was coupled into the input waveguide (left or right arm of the device) by a lensed fiber to generate the −1 or +1 OAM mode through the device. The generated OAM beam was collimated with an objective (100×, NA=0.9), and then split into two branches by a beam splitter. One branch of the OAM beam and the Gaussian reference beam were combined at a beam splitter, and then the interference patterns were captured with an infrared CCD camera (Hamamatsu InGaAs Camera C14041-10U, 320×256 pixels, pixel size 30 μm, high sensitivity in the near infrared region from 950 to 1700 nm, 216 frames/s). The other branch of the OAM beam was converted back to a quasi-Gaussian beam with a vortex hologram loaded onto a spatial light modulator (SLM); it was then coupled into a single mode optical fiber linked with an optical power meter to measure the output power. Here, a half wave plate was used to adjust the polarization of the output OAM mode to match that of the polarization demanded by the SLM. We used a white-light lamp as the illumination source, and a color CCD (Pointgrey Grasshopper GS3-U3-41C6C-C, 2048×2048 pixels, pixel size 5.5 μm, 90 frames/s) was utilized to observe and align the sample.

The intensity distribution was measured by blocking the reference branch in the interferometer. The measured results at 1450, 1500, 1550, 1600, and 1650 nm are shown in Fig. 3 in the main article. The full comparison between the FDTD simulation and the experimental measurements are shown in Fig. S4.

OAM demultiplexing and device alignment

The experimental set-up for the OAM demultiplexing and the alignment between the device and the optical axis of the measurement system is illustrated in Fig. S8. The set-up was similar to the OAM emission system but with a reverse process. The light from a tunable laser passed through a polarization controller and was then collimated into a Gaussian beam by a collimator. After modulation by a vortex hologram loaded onto a SLM, the Gaussian beam was converted into an OAM beam, and its topological charge was determined by the hologram. The OAM beam was focused onto the device by an objective (100× NA=0.9), and its polarization was adjusted with a HWP to match the operation polarization of the device. The output power from both arms was measured through lensed fibers with optical power meters. A white light lamp was utilized as the illumination source and a color CCD was used for the observations.

FDM and OAM (de)multiplexing

A schematic for quadrature phase shift keying (QPSK) and quadrature amplitude modulation (32QAM) optical communication is shown in Fig. S11. The scheme comprises a transmitter system, an OAM emission/sorting and transmission system, and a receiver system. The set-up for the OAM emission/sorting and transmission systems is shown in Figs. S3 and S7; only the transmitter and receiver parts are illustrated.

In the transmitter, a tunable laser (Agilent N7711A, C-band 100 kHz linewidth) was utilized as the pump source. The FDM subcarriers with a frequency separation of 25 GHz (were generated by an optical comb generator (WTAS-02)) were amplified by an erbium-doped fiber amplifier (EDFA); the highly nonlinear fiber broadened the original frequency comb. A programmable optical wavelength selective switch (Finisar WaveShaper 4000S) equalized the power of all 30 optical subcarriers (30-waves), and these subcarriers are then separated into even and odd subcarriers. Both sets of subcarriers were then individually modulated with independent QPSK/32QAM pseudorandom bit sequences. The FDM signal was generated by combining the even and odd subcarriers subsequently. Here the modulation signals were generated with an arbitrary-waveform-generator (Tektronix AWG70002A, sampling rate 25 GS/s), and then amplified with a linear amplifier (SHF 807 with a bandwidth of 30 GHz or SHF S807 with a bandwidth of 55 GHz). These electric signals were transformed to optical signals by a Mach-Zehnder electro-optical modulator. The generated FDM signals were separated into two branches by an optical coupler. One branch was amplified with an EDFA and then directly coupled to the right arm of the device via a lensed fiber to generate the FDM +1 OAM mode. The other branch was delayed with an optical delay line with a 1 ns delay to ensure the signals of the two OAM modes were different; then, the signal was amplified and coupled to the left arm of the device. For the OAM demultiplexing, the two branches of the FDM signals were sent to a SLM to generate the different coaxial OAMs, and the device was utilized to sort these modes and then couple the signals to each arm according to the topological charge of the OAM.

The receiver consisted of a SLM, an EDFA, an optical tunable filter (OTF, Santec OTF-350), a photodiode (PD, U2T XPDV2120R), a linear amplifier (SHF S807, bandwidth 50 GHz), and a real-time oscilloscope with a 50 GSa/s sampling rate (Tektronix DSA72004B). Here, the SLM was utilized for converting the OAM mode beams back to quasi-Gaussian beams and then couple them to a single mode optical fiber. The received signals were amplified with the EDFA, and then each subcarrier could be extracted with the tunable narrow bandpass filter (Santec OTF-350). The light signals were transformed into electrical signals by the photodiode. After amplification with a low noise amplifier, the subcarrier signals were captured with a real time oscilloscope. For the OAM demultiplexing, the SLM was utilized as the OAM mode emitter; thus, the proposed device functions as an OAM sorter.

To quantify the quality of the received signals, bit error rate estimations were performed. The results for the OAM emission are shown in Fig. 4 in the main article, and the bit error rates of the OAMs detection are illustrated in Fig. S14.

**Supplementary Text**

Phase delay for propagating and local resonance

The total phase modulation during the OAM emission process originates both the propagation phase delay and the resonance phase delay from the local subwavelength structure. Thus the phase modulation for the emitter can be expressed as

$$\phi(r,\varphi) = \phi_1(r,\varphi) + \phi_2(r,\varphi), \quad (S1)$$

where $r, \varphi$ are the polar coordinates (origin is the center of the device), $\phi$ is the phase modulation of the emitter, $\phi_1$ and $\phi_2$ are the phase modulations induced by the propagation and localized resonance, respectively. The propagation phase delay is calculated from the integral along the propagation path

$$\phi_1(r,\varphi) = \int_{P(R,\pi)}^{P(r,\varphi)} \frac{2\pi N(r,\varphi)}{\lambda} dl, \quad (S2)$$

where $P(R,\pi)$ is the link point for the device and its left arm, $P(r,\varphi)$, is the local point in the device where the light is emitted out of plane, $N(r,\varphi)$ is the refractive index distribution of the device determined by the detailed substructures. The estimated propagation delay lies somewhere in the range of 0 to 6π, corresponding to a response time less than 15fs. The resonance phase delay is derived with respect to the local refractive index distribution and structure volume as follows:

$$\phi_2(r,\varphi) = \frac{2\pi}{\lambda} N(r,\varphi) c \frac{Q}{\omega_0} + \phi', \quad (S3)$$

where $\tau$ is the group delay caused by the local resonance, $\omega_0$ is the central frequency of the incident beam, $Q$ is the quality factor of the local cavity, $\phi'$ is the difference between the phase delay and the group delay of the local resonance, and its value lies between -π~π.. For a cavity with a volume of $V$, the quality factor can be expressed as

$$Q = \frac{M \omega_0 V}{D}, \quad (S4)$$

where $M$ is the mode density of the cavity, $D$ is the mode degeneracy. For a small cavity, its mode degeneracy can be seen as 1, and the mode density can be written as

$$M = \frac{1}{\omega_0} \frac{8\pi N^3(r,\varphi)}{\lambda^3}. \quad (S5)$$

Thus the quality factor of the cavity can be rewritten as
$$Q = \frac{8\pi N^3(r,\varphi)V}{D\lambda^3}. \tag{S6}$$

We substitute the Eq. (S6) into the Eq. (S3); thus the $\phi_2(r,\varphi)$ is derived as
$$\phi_2(r,\varphi) = \frac{8\pi N^4(r,\varphi)V}{D\lambda^3} + \phi'. \tag{S7}$$

Considering that the substructures of the designed devices have volumes ranging from $1.6\times10^{-4}\lambda^3$ to $4.7\times10^{-3}\lambda^3$. The calculated phase delay is sufficient for a $2\pi$ phase modulation while the corresponding Q of the local cavity is below 5. Therefore, the designed device is equipped with fully phase control and ultrafast time response properties. According to the time-bandwidth product relation, the emitter should also have a broadband spectral domain. We substitute the Eq. (S2) and Eq. (S7) into the Eq. (S1); thus $\phi(r,\varphi)$ is derived as

$$\phi(r,\varphi) = \int_{P(R,\pi)}^{P(r,\varphi)} \frac{2\pi N(r,\varphi)}{\lambda} dl + \frac{8\pi N^4(r,\varphi)V}{D\lambda^3} + \phi'. \tag{S8}$$

Considering that light coming from the left arm yields the $m_1$ order OAM mode, and light coming from the right-hand-side yields the $m_2$ order OAM mode, we obtain the following set of equations:

$$\begin{cases} \int_{P(R,\pi)}^{P(r,\varphi)} \frac{2\pi N(r,\varphi)}{\lambda} dl + \frac{8\pi N^4(r,\varphi)V}{D\lambda^3} + \phi_L' = m_1\varphi; \\ \int_{P(R,0)}^{P(r,\varphi)} \frac{2\pi N(r,\varphi)}{\lambda} dl + \frac{8\pi N^4(r,\varphi)V}{D\lambda^3} + \phi_R' = m_2\varphi. \end{cases} \tag{S9}$$

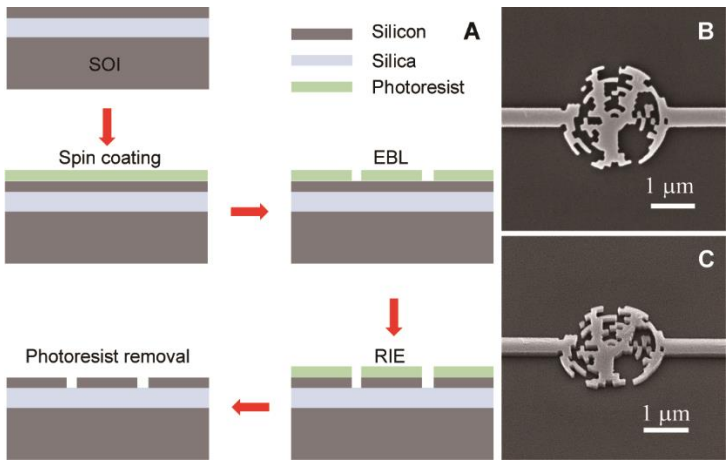

**Fig. S1.**
Fabrication process and scanning electron microscope (SEM) images of the orbital angular momentum (OAM) emitter. (**A**) The device was produced on a standard silicon on insulator wafer, with a 220 nm-thick device layer on top of a 2 μm-thick buried oxide layer. The process flow comprised photoresist spin coating, electron-beam lithography (EBL), deep reactive ion etching (RIE), and photoresist removal. SEM images of the device at two different perspectives: top view (**B**) and 45-degree view (**C**).

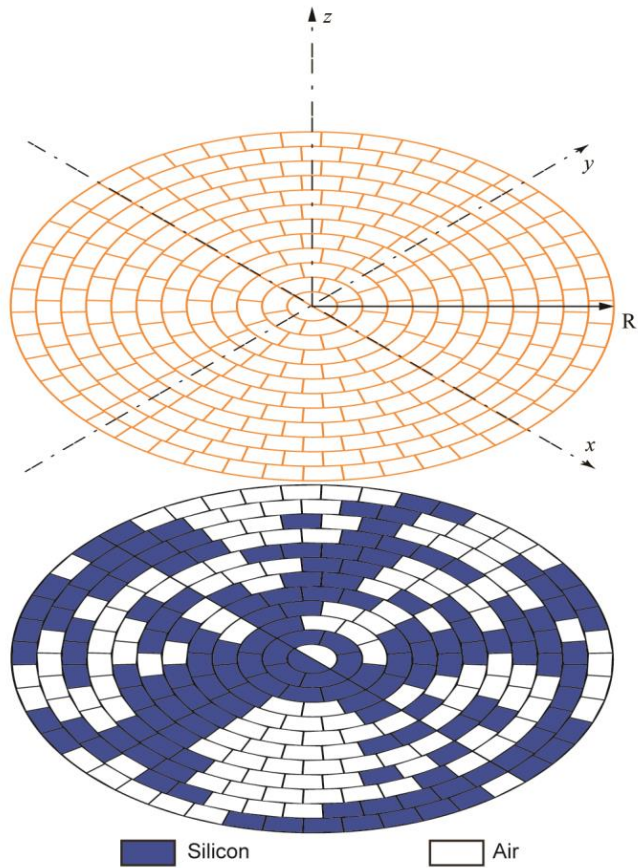

**Fig. S2.**

Substructure design. The device area was circular with a radius of 1.2 µm, and the whole area was divided into 288 subpixels. The intervals in the radial direction were each 100 nm, while the azimuthal interval was also a little less than 100 nm. For each pixel the material was either silicon (blue) or air (white), with the material for each pixel being determined through an optimization algorithm associated with a 3D finite-difference time-domain (FDTD) simulation.

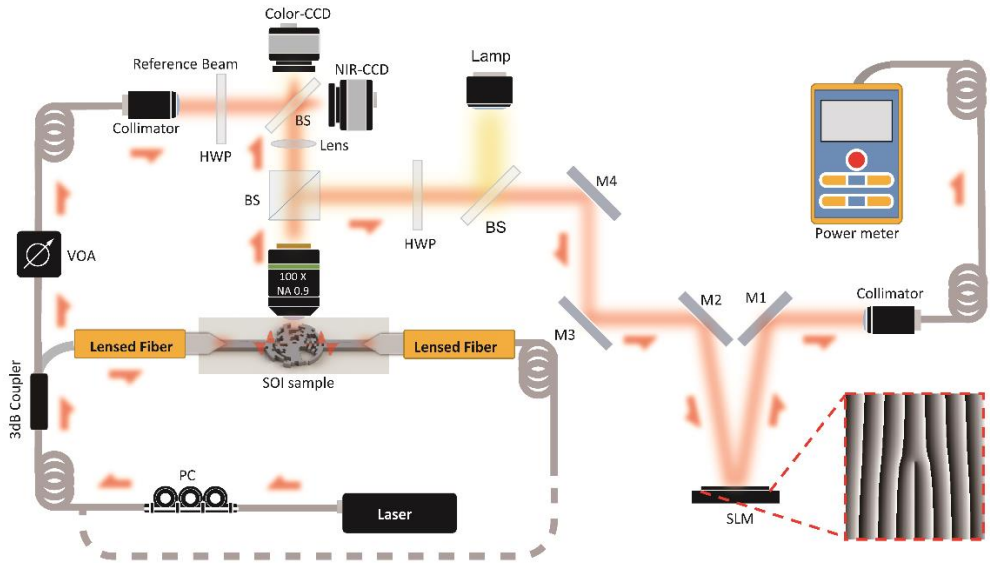

**Fig. S3.**
Experimental setup for characterizing the phase and the intensity profiles of the generated OAM modes from the device and for testing its emission bandwidth. After a polarization controller, the light from a laser that was tunable across the entire telecommunication band (from 1450 to 1650 nm) was split into two branches by a 3-dB optical coupler. One branch was passed to a variable optical attenuator (VOA) (2–60 dB) and to a collimator to produce an expanded Gaussian beam as the reference beam. The other branch was coupled into one of the input waveguides (left/right arm of the device) by a lensed fiber; then either the −1 or +1 OAM mode was generated with the device. The generated OAM beam was collimated with an objective (100×, NA=0.9), and then split into two branches by a beam splitter (BS). One branch of the OAM beam and the Gaussian reference beam were combined at a beam splitter, and then the interference patterns were captured with an infrared CCD camera. The other branch of the OAM was converted back to a quasi-Gaussian beam with a vortex hologram loaded onto the spatial light modulator. The beam was then coupled into a single mode optical fiber linked with an optical power meter to measure the output power. Here, a half wave plate (HWP) was used to adjust the polarization of the output OAM beam to match that of the polarization demanded by the spatial light modulator (SLM).

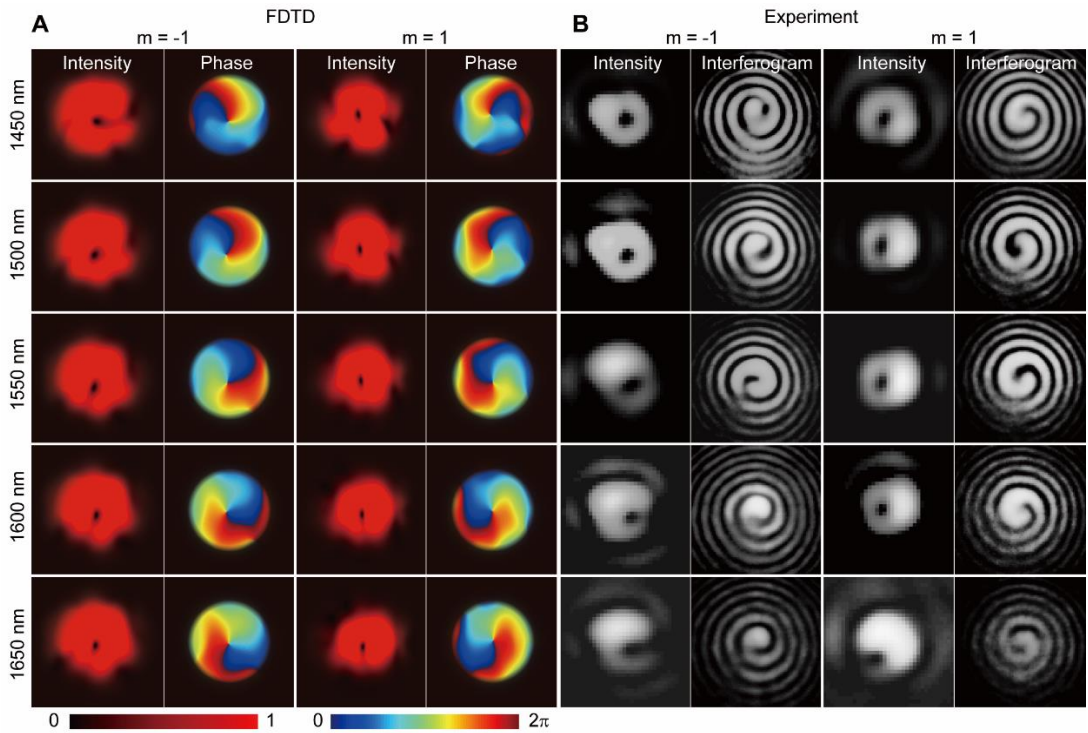

**Fig. S4.**

FDTD simulation and Experimental measurements comparison for the intensity and phase profiles of the generated OAM modes. (**A**) FDTD simulation for near-field intensity and phase distributions for the emission of the −1 and 1 OAM modes at wavelengths of 1450, 1500, 1550, 1600, and 1650 nm. (**B**) Measured far-field intensity distribution and interference patterns for the emission of the −1 and 1 OAM modes at wavelengths of 1450, 1500, 1550, 1600, and 1650 nm.

| Phase distribution on the SLM | Output Efficiency (dB) | |
|---|---|---|
| | Port_1 Input | Port_2 Input |
| Gaussian | -20.2 | -29.7 |
| OV-1 | **-4.60** | -22.2 |
| OV1 | -19.0 | **-5.20** |

**Fig. S5.**

Emission efficiency and mode purity measurement at 1550 nm. The generated OAM was tested using a vortex hologram loaded in the SLM and then coupled into a single mode optical fiber. The output power was measured with an optical power meter and the corresponding efficiency was calculated by taking into account of the insertion loss of SLM, objective lens, beam splitter, and the coupling loss of the lensed fiber.

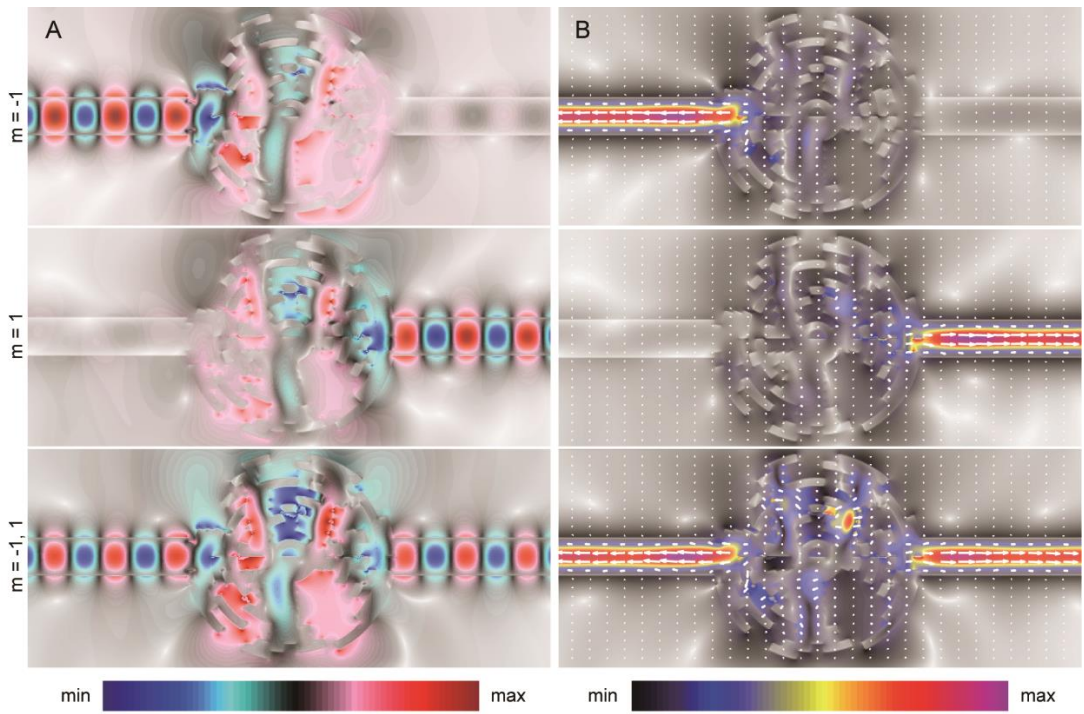

**Fig. S6.**

FDTD simulation for the device as an OAM demultiplexer at 1550 nm for an input beam of OAM m=+1, OAM m=−1, or OAM m=±1. The device couples the −1 mode to its left arm and the +1 mode to its right arm. (**A**) The amplitude distributions for the $E_y$ components. (**B**) The profile for the Poynting vector, where the white arrows indicate the direction of the energy flow.

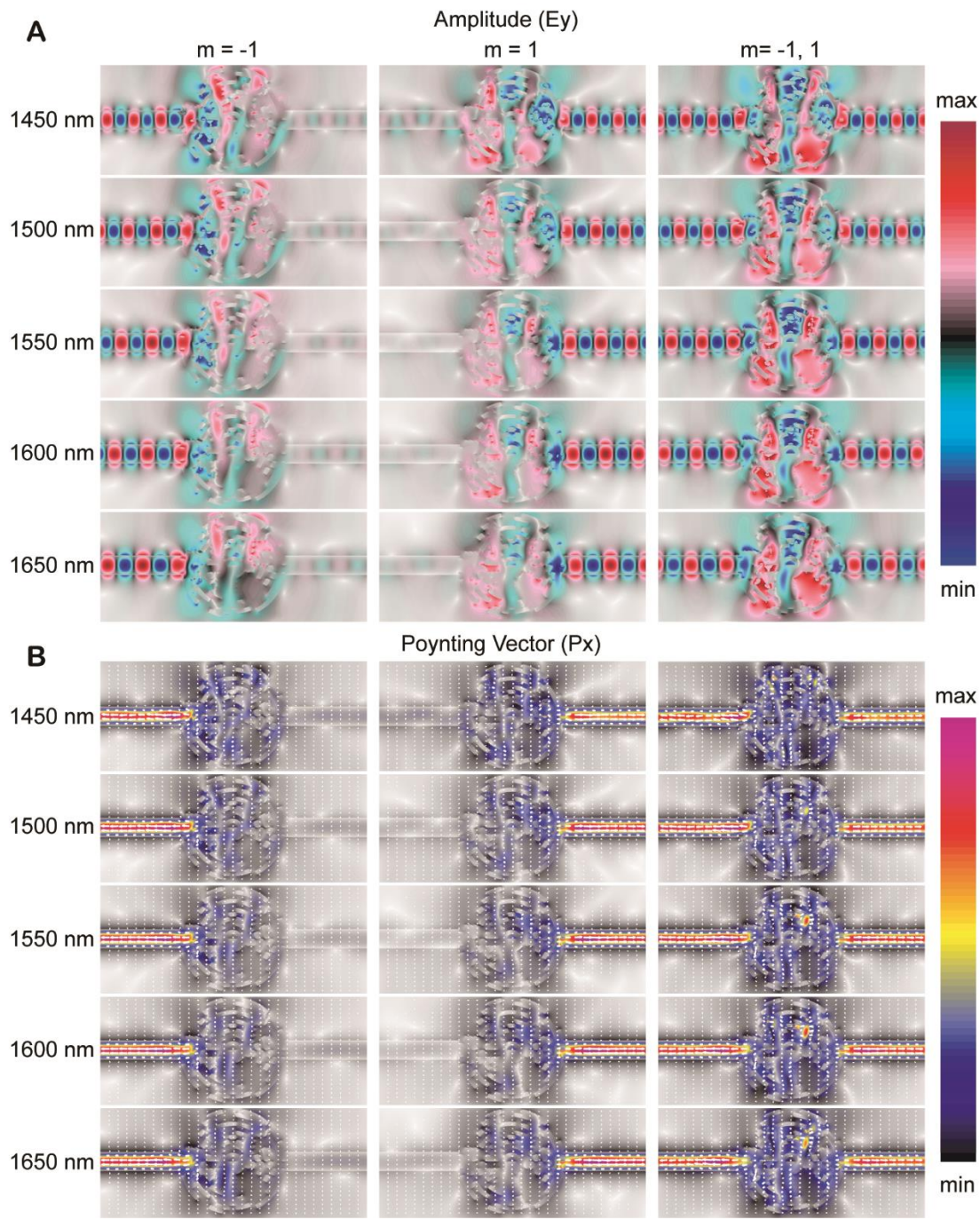

**Fig. S7.**

FDTD simulations for the device as an OAM demultiplexer at 1450, 1500, 1550, 1600, and 1650 nm for an input beam of OAM m=+1, OAM m=−1, or OAMs m=±1. The device works well over the 200 nm-wide regime. (**A**) The amplitude distributions for the $E_y$ components. (**B**) The profile for the Poynting vector, where the white arrows indicate the direction of the energy flow.

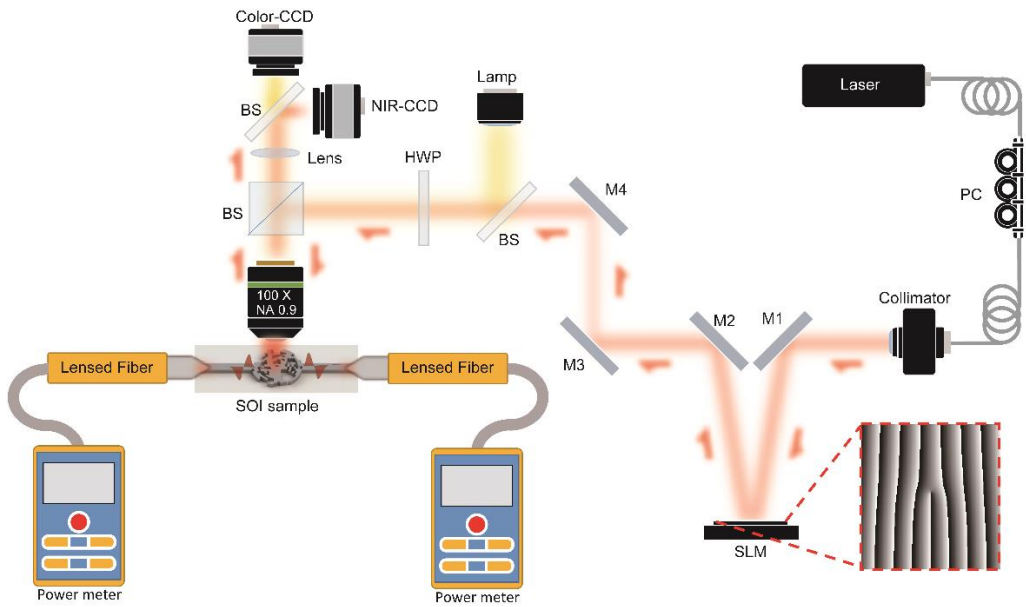

**Fig. S8.**
The experimental set-up for OAM demultiplexing. This set-up was also utilized for the alignment between the device and the optical axis of the measurement system. The light from a tunable laser passed through a polarization controller and was then collimated into a Gaussian beam by a collimator. After being modulated by a SLM loaded with a vortex hologram, the Gaussian beam was converted into an OAM beam, and its topological charge was determined by the hologram loaded on the SLM. The OAM beam was focus onto the device by an objective (100× NA=0.9), and its polarization could be adjusted with a HWP to match the operation polarization of the device. The output power from both arms was measured with optical power meters linked to lensed fiber used to couple the light from the waveguides. A white light lamp was utilized as the illumination source, and a color CCD was used for the observations.

| Input Beam | Output Efficiency (dB) | |
|---|---|---|
| | PORT_1 | PORT_2 |
| Gaussian | -23.5 | -22.7 |
| OV-1 | -6.60 | -33.0 |
| OV1 | -24.4 | -6.80 |

**Fig. S9.**

The efficiency for OAM detection and for the channel segregation measurement for two OAM modes at 1550 nm. The OAM beam is generated with a vortex hologram loaded onto a SLM and then illuminated onto the device. The light was coupled into either the left or right arm of the device depending on the topological charge of the incident OAM beam. The output power was measured with an optical power meter and the corresponding efficiency was calculated by taking into account of the insertion loss of SLM, objective lens, beam splitter, and the coupling loss of the lensed fiber.

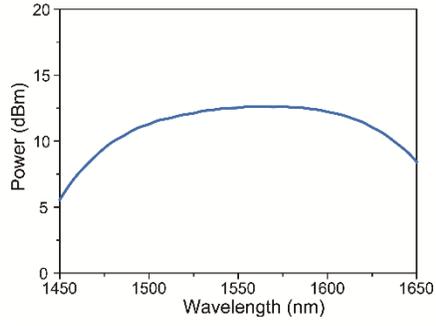

**Fig. S10.**

The light source utilized to characterize the broadband spectrum of the device. The device is a tunable laser with an output wavelength ranging from 1450 to 1650 nm.

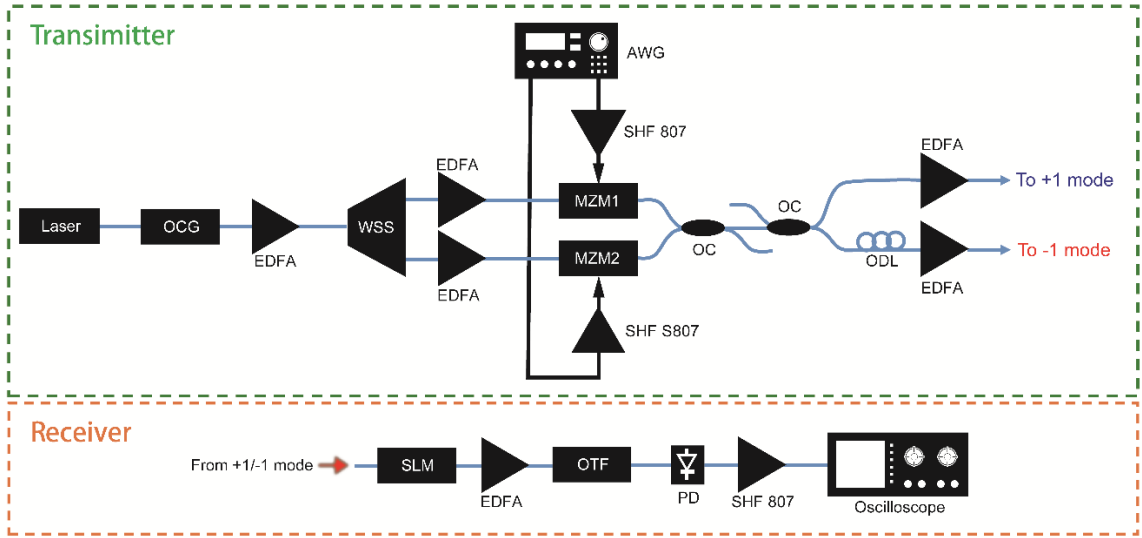

**Fig. S11.**
Schematic for quadrature phase shift keying (QPSK) and quadrature amplitude modulation (32QAM) optical communication. For the transmitter, a tunable laser (Agilent N7711A, C-band 100 kHz linewidth) was utilized as the pump source. The generated frequency division multiplexing (FDM) subcarriers with a frequency separation of 25 GHz from an optical comb generator (WTAS-02) were amplified by an erbium-doped fiber amplifier (EDFA); the highly nonlinear fiber broadened the original frequency comb. A programmable optical wavelength selective switch (Finisar WaveShaper 4000S) equalizes the power of all the optical subcarriers (30-waves), and these subcarriers were then separated into even and odd subcarriers. Both sets of subcarriers were then individually modulated with independent QPSK/32QAM pseudorandom bit sequences. The FDM signal was subsequently generated by combining the even and odd subcarriers. Here the modulation signals were generated with an arbitrary-waveform-generator (Tektronix AWG70002A, sampling rate 25GS/s) and then amplified with a linear amplifier (SHF 807 with a bandwidth of 30 GHz or SHF S807 with a bandwidth of 55 GHz). These electric signals were transformed into optical signals using a Mach-Zehnder electro-optical modulator (MZM). The generated FDM signals were separated into two branches by an optical coupler. One branch was amplified with an EDFA and was then directly coupled to the right arm of the device by a lensed fiber to generate the FDM +1 OAM mode. The other branch was delayed with an optical delay line (ODL) with a 1 ns delay to ensure the signals of the two OAM modes were different; then the signal was amplified and coupled to the left arm of the device. The receiver comprised a spatial light modulator (SLM), an optical tunable filter (OTF, Santec OTF-350), a photodiode (PD, U2T XPDV2120R), a linear amplifier (SHF S807, bandwidth 50 GHz), and a real-time oscilloscope with a 50 GSa/s sampling rate (Tektronix DSA72004B).

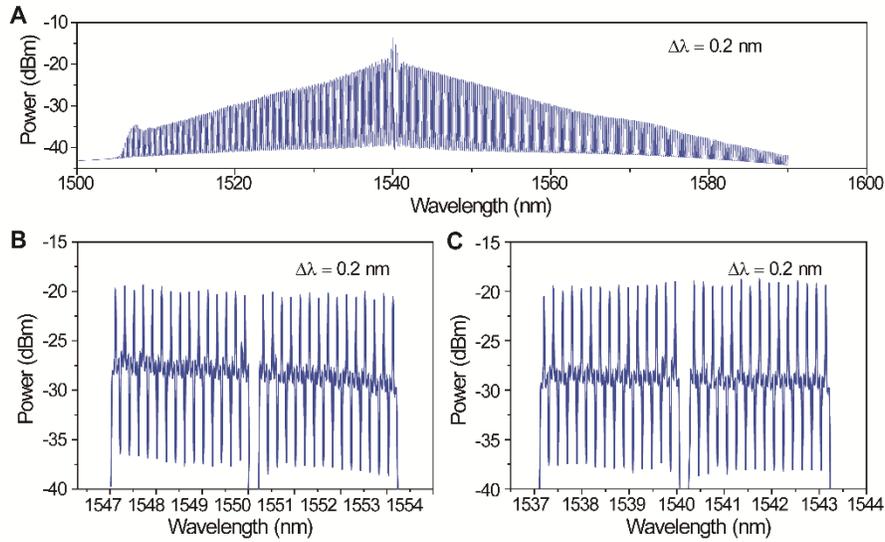

**Fig. S12.**
(**A**) Femtosecond optical frequency comb from an optical frequency comb generator. (**B**) The 30-wave FDM signal (from 1547 to 1554 nm, wavelength interval of 0.2 nm) for OAM mode multiplexing communication (here the emitter was used as a transmitter). These 30-wave channels were filtered out from the femtosecond optical frequency comb with a wave shaper. (**C**) The 30-wave FDM signal (from 1537 to 1544 nm, wavelength interval of 0.2 nm) for OAM mode demultiplexing communication (the emitter was used as a receiver).

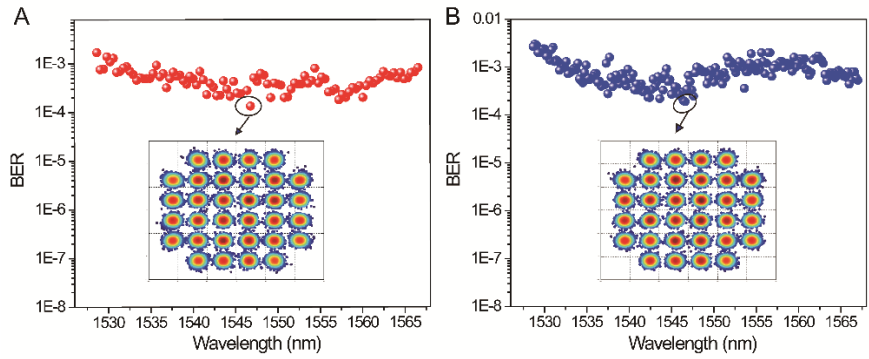

**Fig. S13**

32QAM communication experiment results with wavelength scanning from 1527 to 1567 nm with a wavelength scanning interval of 0.2 nm. (**A**) Bit error rate (BER) measurement for the +1 OAM mode generation. (**B**) BER measurement for the +1 OAM mode detection.

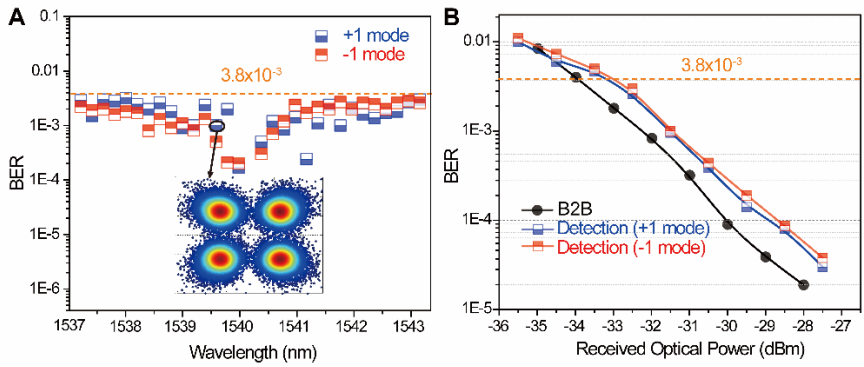

**Fig. S14**

QPSK communication results for 30-wave and two OAM mode demultiplexing (the device was used as a receiver). (**A**) The measured BER of the +1 OAM mode (blue rectangles) and −1 OAM (red rectangles). The BERs of all channels were below the hard decision forward error correction limit of $3.8 \times 10^{-3}$. The insert shows the corresponding signal constellations when the BER is $1.0 \times 10^{-3}$. (**B**) BER measurements for back-to-back (B2B) test case, +1 and -1 modes detection at 1540 nm.